\documentclass[twocolumn,english]{revtex4-1}
\usepackage[T1]{fontenc}
\usepackage[latin9]{inputenc}
\usepackage{geometry}
\usepackage{graphicx,color}

\makeatletter
\@ifundefined{date}{}{\date{}}
\makeatother

\usepackage{babel}
\begin{document}
\title{Loss of material trainability through an unusual transition}
\author{Himangsu Bhaumik}
\author{Daniel Hexner}
\email{danielhe@me.technion.ac.il}
\affiliation{Faculty of Mechanical Engineering, Technion, 320000 Haifa, Israel}
\begin{abstract}
Material training is a method to endow materials with specific responses
through 
external driving. We study the complexity of attainable responses, as expressed in the
number of sites that are simultaneously controlled. With
increased complexity, convergence to the desired response becomes very slow. The training error decays as a power-law with an exponent that varies continuously and vanishes at a critical threshold, marking the limit of trainable responses. We study how the transition affects the vibrational properties. Approaching the critical threshold, low frequency modes proliferate, approaching zero frequency. This implies that training causes material degradation and that training fails due to competing spurious low frequency modes. We propose that the excess low frequency spectrum is due to atypical local structures with bonds that nearly align. Our work explains
how the presence of an exotic critical point affects the convergence
of training, and could be relevant for understanding learning in physical
systems.
\end{abstract}

\maketitle
Networks are a common representation of natural and engineered systems
that process high dimensional set of inputs\citep{barzel2013universality}.
These include, natural and synthetic neural networks\citep{bassett2017network,kotsiantis2007supervised,mehta2019high}
that perform computations or store memories\citep{hopfield1982neural},
regulatory networks\citep{davidson2010regulatory} and more recently,
elastic and flow networks that encode complex responses\citep{mitchell2016strain,rocks2017designing,yan2017architecture,rocks2019limits,hexner2020periodic}.
A central challenge in these system is understanding the capacity
of such networks, in terms of the complexity of the responses that
can be attained. It is not surprising that as the difficulty of the
response is increased, the system fails to yield the desired behavior.
These limitation can be either attributed to the capacity of the network
itself or the algorithm by which the parameters are adjusted\citep{schoning1999probabilistic}.

In this paper we consider the capacity of mechanical networks that
are trained through sequences of applied strains. We build on recent
work that demonstrated that disordered networks can be trained with externally applied fields\citep{hexner2020periodic,pashine2019directed}.
During the training the external driving generates stresses
which cause the network to remodel its structure through plastic deformations.
As a result the system evolves to yield the desired response. 
Training can be considered a learning process, where the material itself learns without the use of a computer\citep{scellier2017equilibrium,stern2020supervised2,kendall2020training,stern2021supervised}.
In contrast to designed structures, this approach does not require
fabrication, nor manipulating directly the microscopic structure,
and is therefore potentially scalable.

We focus on the convergence of the response as a function of the number
of target sites that are simultaneously controlled. As the number of target sites
is increased training becomes very sluggish. The training error decays approximately as
a power-law with an exponent that varies continuously with the number
of trained sites. At a critical threshold the exponent appears to
vanish indicating a phase transition. Unlike conventional critical
points, here, power-laws are observed over a broad range of parameters
\citep{noest1986new,moreira1996critical,janssen1997renormalized,dickman1998violation,vojta2006rare}.
The transition also marks the limits of attainable responses. We find
that the capacity, defined by the number of sites that can be simultaneously
controlled, is approximately extensive, scaling with system size.

We conclude by searching for a signature of this transition in the
vibrational properties. We find a substantial
increase in the low frequency spectrum as complexity is increased.
The trained response can be associated with a single low frequency
mode that is separated from the remaining of the spectrum\citep{hexner2021adaptable}.
Here, failure is accompanied with a proliferation of low frequency
modes that compete with the desired response. At criticality the density of states appears to creep down to arbitrarily small frequencies.  We provide evidence that the excess low frequencies modes are due to non-generic geometries, characterized by bonds that align.

\noindent\textbf{\emph{Model:}} We employ the model and training protocol of
Ref. \citep{hexner2020periodic}, which is briefly summarized. We
model an amorphous material as a network of springs in two dimensions. The force on
each spring is given by $k_{i}\left(\ell_{i}-\ell_{i,0}\right)$,
where $\ell_{i}$ is the length of the bond, $\ell_{i,0}$ the rest
length and $k_{i}$ is the spring constant. For convenience we consider
networks that are derived from amorphous packing of repulsive spheres
at zero temperature. The networks are characterized
by their coordination number $Z=\frac{2N_{b}}{N}$, where $N_b$ is the number of bonds and $N$ is the number of nodes. Rigidity requires that $Z>Z_{c}\simeq2d$\citep{maxwell1864calculation,CALLADINE,PELLEGRINO1,alexander1998amorphous}.

\begin{figure}
\begin{centering}
\includegraphics[width=1\linewidth]{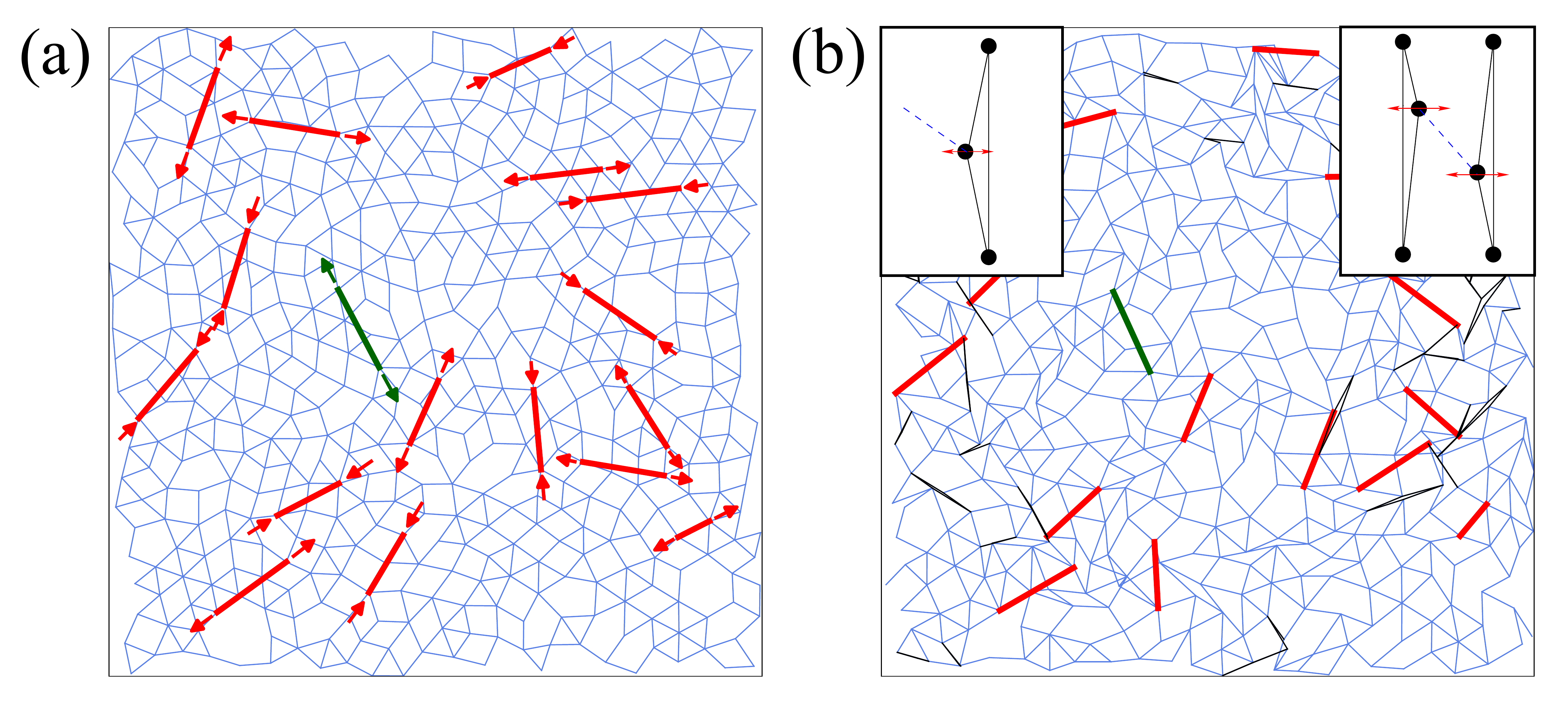}
\par\end{centering}
\caption{An example of (a) an initial network and (b) a trained network (with $N=500$ nodes). Each pair of the source sites and target sites are marked by a connecting green and red line respectively. The response of the targets can either be in-phase or out-of-phase with respect to the source, as indicated by the arrows. In the trained network the angles between adjacent bonds may become small (bonds with $\theta<10^{o}$ are drawn in black). The left inset shows a  structural motif which has a soft direction  (marked in red) when detached from the network by cutting the blue bond. The right inset shows two coupled motifs that contribute to a localized low frequency modes. \label{fig:Fig0}}
\end{figure}

Our goal is it to train responses where a single input strain at a source
site yields a prescribed strain on $N_{T}$ randomly chosen target
sites, as illustrated in Fig. \ref{fig:Fig0}. Each source and target
sites are pairs of nearby nodes and the local strain is defined as
the fractional change in their distance. We denote that strain on
the source and target by $\epsilon_{S}$ and $\epsilon_{T}$. 
The strain on both the source and targets is
chosen to have the same amplitude, $\epsilon_{Age}$, however, the response is chosen to be in-phase or out-of-phase with equal probability.

Training relies on plastic deformations that alter the structure of the network. Here, we only consider changes to the rest lengths \citep{hexner2020effect}.
Each bond is modeled as a Maxwell viscoelastic
element \citep{maxwell1867iv}, where the change in the rest length is
proportional to the force on the bond:
\begin{equation}
\partial_{t}\ell_{i,0}\propto k_{i}\left(\ell_{i}-\ell_{i,0}\right).\label{eq:Plastic}
\end{equation}
We focus on the quasistatic regime, where the time to reach force
balance is small with respect to the time scales of plasticity.
In simulations time is discritized in to small steps, where at each
step, we vary the strain, minimize the energy to reach force balance,
and then evolve the rest length in accordance with Eq. \ref{eq:Plastic}.

The response of an elastic network is dominated by the softest direction in the energy landscape, and therefore we aim at sculpting an energy
``valley'' that couples the input source site and an output target
sites. In our model, the rest lengths evolve to lower the internal
stresses and elastic energy. Training an energy valley is therefore  performed by cyclically straining the source and targets along the desired response, while allowing plastic deformation sculpt the energy landscape.  The source and target sites are strained by attaching the pairs of nodes with ``ghost bonds'' and varying their rest length.

\noindent\textbf{\emph{Convergence \& Phase Transition:}} To test the generality
of our results we consider both small and large $\Delta Z\equiv Z-Z_{c}$, with the corresponding value of $\Delta Z\approx 0.03$ and $\Delta Z\approx 0.75$ respectively.
In the small $\Delta Z$ limit the networks are nearly isostatic,
and elasticity is anomalously long ranged \citep{ellenbroek2006critical,lerner2014breakdown}.
The long range response to pinching a bond has been shown to be useful
in coupling distant sites during training \citep{hexner2020periodic}.
Therefore, for large $\Delta Z$ training is only successful for sufficient
number of targets\citep{hexner2020periodic}. Nonetheless, we find
that the qualitative results are similar.

\begin{figure}
\begin{centering}
\includegraphics[width=1\linewidth]{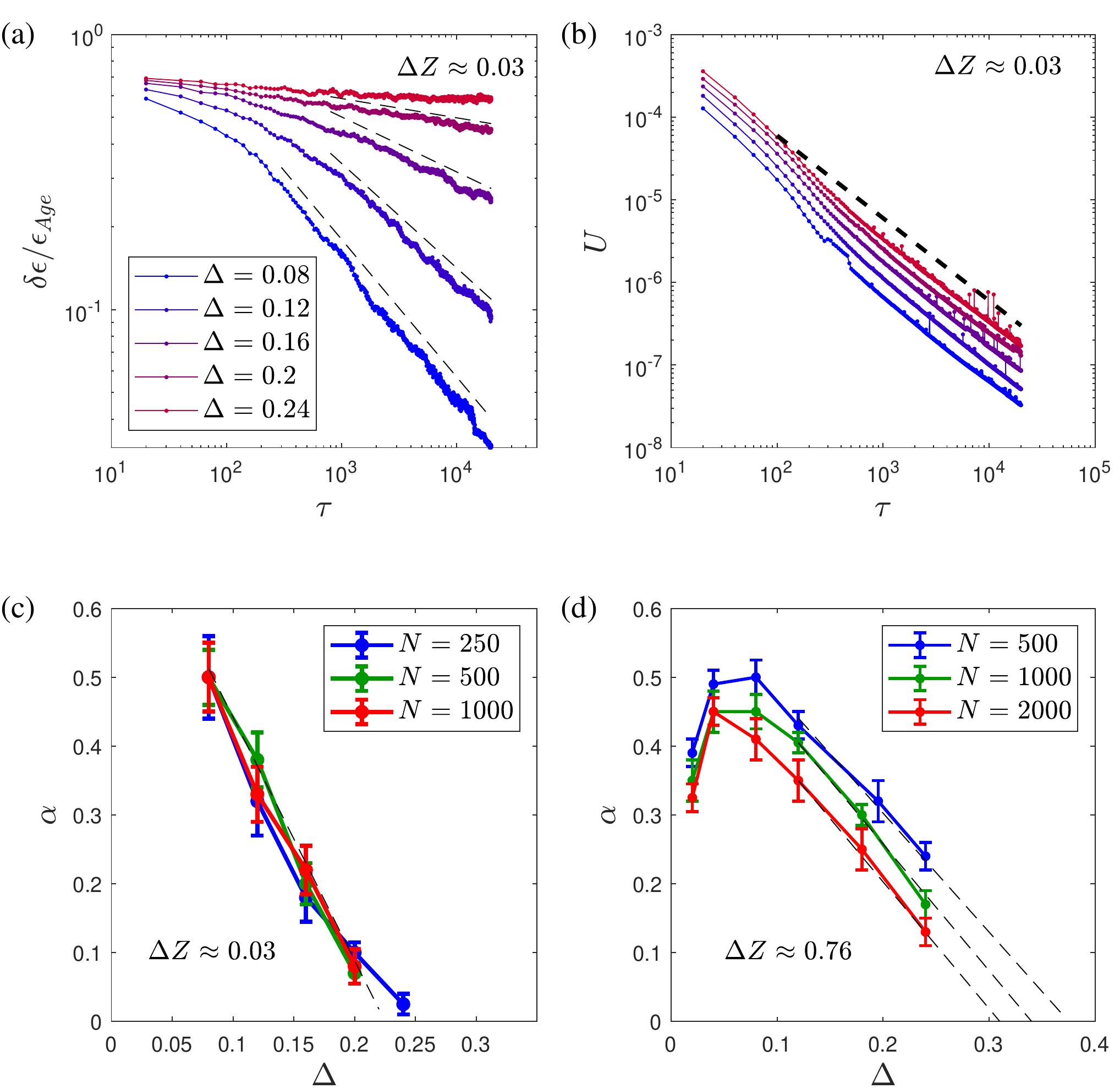}
\par\end{centering}
\caption{Characterizing convergence as a function of the number of targets
per node, $\Delta=N_{T}/N$. (a) The training error as a function
of the number of cycles. With increased $\Delta$, convergence becomes
very slow, decaying approximately as a power-law. The dashed lines
are a guide to the eye. (b) The decrease in energy along the training
path is weakly dependent on $N_{T}$. The exponent $\alpha$ as a function of $\Delta$ for small (c)  and large (d) coordination number.
$\alpha$ appears to vanish at a critical threshold that depends weakly on $N$.  In (a)(b)
$N=500$.\label{fig:Fig1}}
\end{figure}

We characterize the convergence of the response by measuring the error,
$\delta\epsilon$, defined as the difference in absolute value from
the desired response. We average the error over an entire cycle, where
the strain is varied up to $\epsilon_{Age}$, and over the number
of targets. Fig. \ref{fig:Fig1}(a) shows $\delta\epsilon/\epsilon_{Age}$
as a function of the number of training cycles for different number
of target sites, $N_{T}$, as conveniently indicated by $\Delta\equiv\frac{N_{T}}{N}$.
As $\Delta$ increases the convergence becomes slower and slower.
At long times $\delta\epsilon$ decays approximately as a power-law,
\begin{equation}
\delta\epsilon\propto\tau^{-\alpha}.
\end{equation}
Fig. \ref{fig:Fig1}(c) shows that $\alpha$ depends on $\Delta$;
it decreases with $\Delta$ and appears to vanish at a finite value $\Delta_{c}$. Exploring the regime near $\Delta_{c}$ is
unavoidable difficult due to the slow convergence rate.

We interpret the point $\Delta=\Delta_{c}$ as a critical point separating
converging responses, from unattainable responses. This transition
is unlike conventional continuous phase transitions where power-law
scaling occurs only at the critical point. Here, over the entire range
of $\Delta<\Delta_{c}$ convergence scales as a power-law, which implies
extremely slow convergence. Assuming we are satisfied with an error,
$\delta\epsilon_{m}$, the time required scales as $\tau\propto\left(\delta\epsilon_{m}\right)^{-1/\alpha}$.
Taking, $\alpha\propto\left|\Delta-\Delta_{c}\right|^{\beta\approx1}$
yields a Vogel-Fulcher-Tammann like law\citep{VFT1,VFT2,VFT3} $\tau\propto e^{A\left|\Delta-\Delta_{c}\right|^{-\beta}}$,
where $A=-log\epsilon_{m}$. This is far slower than the power-law
divergence in conventional phase transitions $\tau\propto\left|\Delta-\Delta_{c}\right|^{-\theta}$.

We note that similar behavior has been observed in another class of
non-equilibrium phase transitions, that separates a static absorbing
phase from a chaotic phase\citep{grassberger1979reggeon,hinrichsen2000non}.
The directed percolation class has the characteristics of a critical
transition with power-law scalings, however, when quenched disordered
is added the nature of the transition changes and there is a regime
where activity decays as a power-law with a continuously varying exponent
\citep{noest1986new,moreira1996critical,janssen1997renormalized,dickman1998violation,vojta2006rare}.
The effect where disorder alters the pristine transition is known
as a Griffiths phase. The origin of this behavior can be traced to
a rare regions that have an overwhelming contribution.

In our system, there is also a convergence to absorbing states. Training
reduces the energy along a prescribed path until the energy vanishes.
At that point, when there are no longer any internal forces the system
ceases to evolve, thus, reaching an absorbing state. However, in contrast
to the strong dependence of the error on $\Delta$, the decay of the
elastic energy shown in \ref{fig:Fig1}(b) is nearly independent of
$\Delta$ and is relatively quick, scaling approximately as $\tau^{-1}$.
This suggests that regardless of $\Delta$ the evolution of the networks
slows down and ultimately freezes. 

\textbf{Capacity: }Next, we consider the system size dependence. Fig
\ref{fig:Fig1}(c) shows that for small $\Delta Z$ the exponent $\alpha$
is very weakly dependent on system size. This suggests that 
$\Delta_{c}$ is approximately constant, implying the number of sites that
can be simultaneously trained is extensive, proportional to $N$.
Fig. \ref{fig:Fig1}(d) shows $\alpha$ as a function of $\Delta$
for systems with large $\Delta Z$ and for different system sizes. The exponent, $\alpha$, is maximal at an intermediate value of $\Delta$; there, it depends weakly on system size. We estimate $\Delta_{c}\left(N\right)$ by extrapolating to the point where $\alpha$ vanishes and find that it slightly decreases with $N$. Over a four fold increase in system size, $\Delta_{c}$
changes only by approximately $13\%$. Thus, capacity is nearly extensive.
Despite this slight decrease of capacity with system size, the large
$\Delta Z$ networks have a larger $\Delta_{c}$ than of the small
$\Delta Z$ networks.

We note that the (near) extensive capacity is different than the sub-extensive
scaling found in tuning networks by bond removal and addition in Ref.
\citep{rocks2019limits}. 

Lastly, we note that there is an additional system dependent time
scale \footnote{see supplementary information}, which marks the crossover to the power-law regime. We find
that this time scale grows approximately as $N^{\approx0.6}$, implying
that larger systems take longer time to train (see Supplementary Information S-4). We rationalize this time
scale by noting that training is only successful when the stiffness
along the training path falls below the remaining transverse stiffnesses, which are defined by the system's eigen-frequencies. The lowest frequency in the system on average decreases with system size and therefore larger systems require more training to further reduce the stiffness along the trained path.

\noindent \textbf{\emph{Normal mode analysis: }}As noted, the response depends
on the ratio of the stiffness in the transverse directions to the stiffness along the  trained path.  Since the decay of energy in Fig. \ref{fig:Fig1}(b)
weakly depends on $\Delta$, this suggests that transverse stiffness
becomes small. Within  linear response, the stiffness corresponds to the eigen-frequency, which are characterized by the density of states. To this end, we compute the Hessian, $H$, defined by the matrix of second derivatives of the energy, and diagonalize it to find its spectrum. Previously it was found, that the lowest mode corresponds to the trained response\citep{hexner2021adaptable}, and therefore the remaining frequencies correspond to the transverse stiffnesses.

\begin{figure}
\includegraphics[width=1\linewidth]{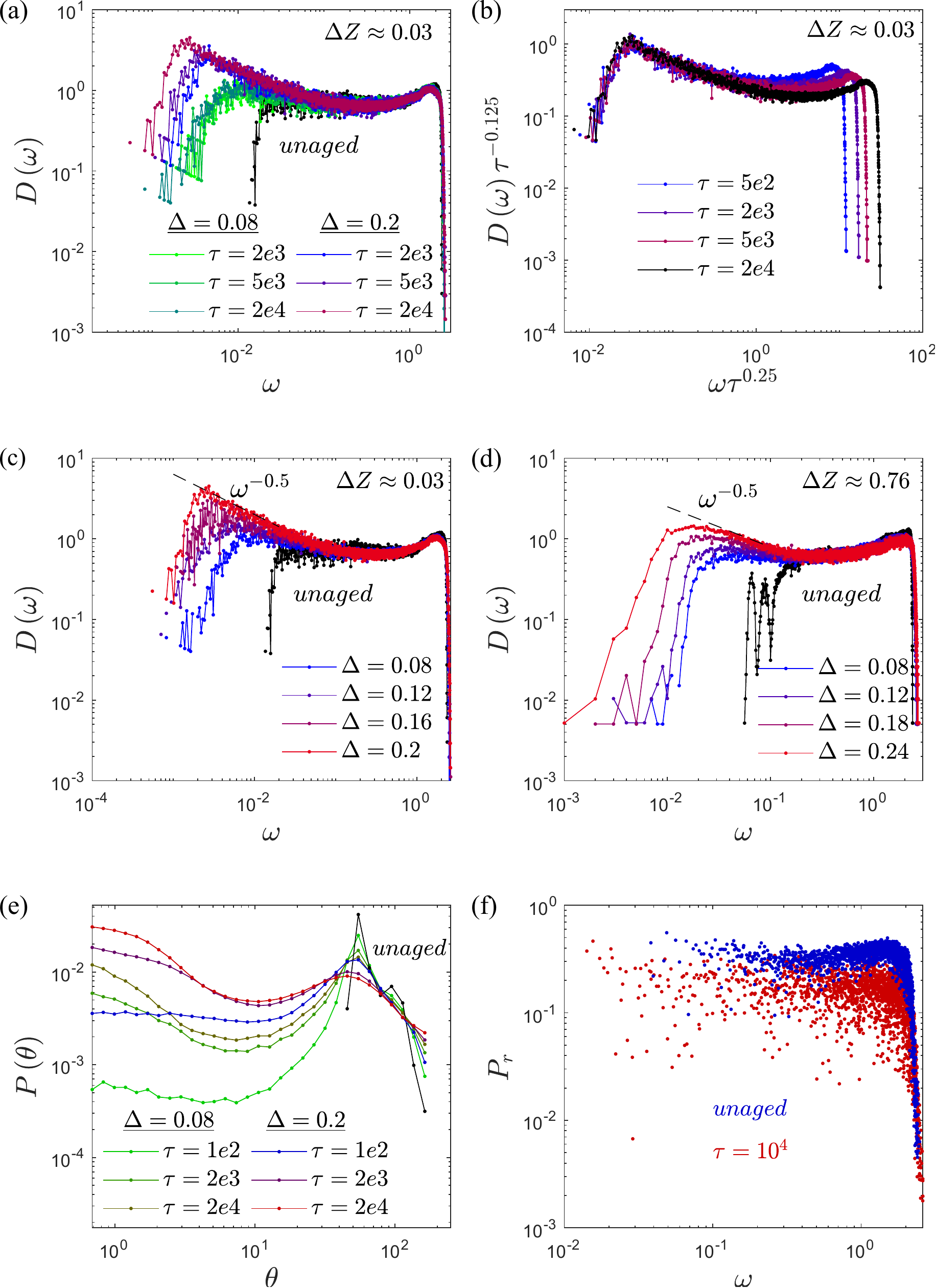}

\caption{Characterization of the vibrational properties. (a) The evolution of the
density of states with the number of cycles. For small $\Delta$ the
density of states ceases to evolve after $\tau\sim200$ cycles, while
for larger $\Delta$ it continuously shifts to lower frequencies with
additional training. (b) A collapse of the density of states near
the transition ($\Delta=0.2$), suggests that the shift to lower frequencies
scales as $\omega_{c}\propto\tau^{-0.25}$. (c) and (d) The density
of state at a fixed number of cycles, $\tau=2\times10^{4}$ for different
values of $\Delta$. In (c) $\Delta Z\approx0.03$, while in (d) $\Delta Z\approx0.76$.
In both cases the shift to lower frequencies grows with $\Delta$. (e) The distribution of the angles (measured in degrees) between bonds for different number of cycles. (f) The participation ratio before training (blue) and after $10^4$ training cycles (red). 
In (e),(f) $N=1000$, $\Delta Z\approx0.76$. \label{fig:DOS}}
\end{figure}

Prior to training there are very few modes below a characteristic
frequency, $\omega^{*}\propto\Delta Z$ \citep{Ohern}. Fig. \ref{fig:DOS}(a)
shows the evolution of the density of states, $D\left(\omega \right)$, for two values of $\Delta$.
For clarity, we exclude the lowest frequency mode associated with
the trained response. We observe two behaviors depending on the distance
to $\Delta_{c}$. Near $\Delta_{c}$ there is a continual shift towards
lower frequencies with the number of cycles. Whereas, away from $\Delta_{c}$
the density of states is nearly stationary after about $\sim200$
cycles. To probe the evolution of the density of states near the
transition we collapse the curves at low frequencies. Fig. \ref{fig:DOS}(b) shows that shift to lower frequencies scales  approximately as $\tau^{-0.25}$, suggesting a continual shift to zero frequency.  This feature is not present in previously found Griffiths phases.

Next, we study the dependence of the density of states on $\Delta$ at fixed $\tau$. Fig. \ref{fig:DOS}(c) and (d) show the density of states for small and large $\Delta Z$ correspondingly. In both cases,
as $\Delta$ increases there is a shift towards lower frequencies.
The trend is suggestive that asymptotically, at $\Delta_{c}$ the
density of states extends to zero frequency. At small frequencies, $D \left(\omega \right)$ can be fitted by $\omega^{-0.5}$.  

The enhanced low frequency spectrum explains how training fails. As noted, the response is governed by the
ratio of the stiffness along the trained direction and the
transverse stiffnesses. The proliferation of low frequency modes indicates that there are many competing spurious modes. Therefore, more training cycles are needed to decrease the stiffness along the training path. In the supplementary information we show the error scales as the ratio of the two lowest frequencies squared. We also note, that retraining a material repeatedly also results in an excess in the low frequency spectrum\citep{hexner2021adaptable}.

\noindent \textbf{\emph{Structural features \& Nature of low frequency modes:}} Next, we search for the origin of the excess low frequency modes in the structure.  Fig. \ref{fig:Fig0} shows the network before (a)  and after training (b). In the trained network some of the angles between bonds becomes very small. Prior to training the smallest angle  formed by adjacent bonds is, $\theta\approx39^o$. As shown in Fig. \ref{fig:DOS}(e) the distribution at small angle becomes pronounced with increased training, especially for large $\Delta$.  To relate the small angles to the low frequency modes, we note that an unstressed bond assigns an energy cost for node displacements that alter the bond's length. This can be considered a constraint on the zero energy motions. When two bonds align ($\theta= 0,\pi$) these two constraints are redundant, reducing the stiffness along the transverse direction.

We have identified local structural motifs where bond alignment leads to soft deformations. We illustrate this in the example shown in the inset of Fig. \ref{fig:Fig0}(b) (left). In the absence of the bond by which the motif is attached to the network (marked in blue) the transverse motion is soft, but constrained when the bond is present. However, two motifs can couple  (see inset (right)) to form a soft energy deformation. If the bonds in the motifs perfectly align this leads to a local zero mode. Motif where all the bonds of a given node align yield a soft deformation, however this motif is rare. Thus, through individual and coupled motifs the eigen-frequencies are shifted to lower values. 
 
We also characterize the low frequency modes by computing the participation ratio $P_r$ \footnote{See the supplementary information for additional details}. As shown in Fig. \ref{fig:DOS}(f), $P_r$ of the trained network decreases over the whole frequency range. We associate the increased localization with local soft motifs. In addition, and similarly to the untrained network, there are also extended modes. Additional discussion on the soft motifs and normal modes is provided in the supplementary information.
  
 We consider the enhancement of the low frequency spectrum as degradation.
The shift towards lower frequencies implies that there is an overall softening
of the system. In the supplementary information we show that both the bulk and shear modulus decrease with the number of training cycles, in particularly near $\Delta_c$. We therefore believe
that this transition could be of interest in studying aging under periodic drive (or material fatigue).

\noindent\textbf{\emph{Conclusions:}} In summary, we have studied the effect
of varying the complexity of the trained responses. The most prominent
behavior is that convergence becomes very slow with increasing difficulty,
manifested through a power-law decay of the error. At a critical threshold the exponent vanishes, implying that the convergence time diverges with a Vogel-Fulcher-Tammann like law. At small $\Delta Z$ we find that the critical threshold $\Delta_{c}$ is independent of system size, implying that responses with an extensive number of sites can be trained. At larger connectivity the capacity appears nearly extensive, with a weak system size dependence.

To characterize the transition we also studied the density of states and the normal modes. With increased difficulty, the
density of states creeps towards lower frequencies. Near the transition it appears that the low frequency modes reach arbitrarily small frequencies with sufficient training. We show that the growing number
of lower energy modes compete with the desired response. The system
cannot distinguish between the trained response and the spurious low
energy modes. We have also studied the structural source of the low frequency modes and characterized the eigen-modes. 

Our work demonstrates how an exotic critical point affects training.
The nearness to the critical point defines a difficulty measure, and
while we have focused on the number of targets, the difficulty of
the trained response could have other contributions, including, the
non-linearity of the response\citep{hexner2020training} or its strain
amplitude. With increased difficulty, there is also an increase in
degradation, marked by the enhanced low frequency spectrum. This could
affect robustness of the response as well as the number of times the
system can be retrained\citep{hexner2021adaptable}. It is interesting
to ask how universal this type of transition is, and whether it applies to other forms of training. In particular, since training is intimately related to learning\citep{scellier2017equilibrium,stern2020supervised2,kendall2020training,stern2021supervised}, perhaps a similar transition occurs in other learning algorithms.
\begin{acknowledgments}
We would like to thank Dov Levine and Andrea J. Liu
for enlightening discussions. This work was supported by the Israel
Science Foundation (grant 2385/20) and the Alon Fellowship.
\end{acknowledgments}

\bibliographystyle{apsrev4-2}
\addcontentsline{toc}{section}{\refname}\bibliography{biblo}

\begin{thebibliography}{40}%
\makeatletter
\providecommand \@ifxundefined [1]{%
 \@ifx{#1\undefined}
}%
\providecommand \@ifnum [1]{%
 \ifnum #1\expandafter \@firstoftwo
 \else \expandafter \@secondoftwo
 \fi
}%
\providecommand \@ifx [1]{%
 \ifx #1\expandafter \@firstoftwo
 \else \expandafter \@secondoftwo
 \fi
}%
\providecommand \natexlab [1]{#1}%
\providecommand \enquote  [1]{``#1''}%
\providecommand \bibnamefont  [1]{#1}%
\providecommand \bibfnamefont [1]{#1}%
\providecommand \citenamefont [1]{#1}%
\providecommand \href@noop [0]{\@secondoftwo}%
\providecommand \href [0]{\begingroup \@sanitize@url \@href}%
\providecommand \@href[1]{\@@startlink{#1}\@@href}%
\providecommand \@@href[1]{\endgroup#1\@@endlink}%
\providecommand \@sanitize@url [0]{\catcode `\\12\catcode `\$12\catcode
  `\&12\catcode `\#12\catcode `\^12\catcode `\_12\catcode `\%12\relax}%
\providecommand \@@startlink[1]{}%
\providecommand \@@endlink[0]{}%
\providecommand \url  [0]{\begingroup\@sanitize@url \@url }%
\providecommand \@url [1]{\endgroup\@href {#1}{\urlprefix }}%
\providecommand \urlprefix  [0]{URL }%
\providecommand \Eprint [0]{\href }%
\providecommand \doibase [0]{https://doi.org/}%
\providecommand \selectlanguage [0]{\@gobble}%
\providecommand \bibinfo  [0]{\@secondoftwo}%
\providecommand \bibfield  [0]{\@secondoftwo}%
\providecommand \translation [1]{[#1]}%
\providecommand \BibitemOpen [0]{}%
\providecommand \bibitemStop [0]{}%
\providecommand \bibitemNoStop [0]{.\EOS\space}%
\providecommand \EOS [0]{\spacefactor3000\relax}%
\providecommand \BibitemShut  [1]{\csname bibitem#1\endcsname}%
\let\auto@bib@innerbib\@empty
\bibitem [{\citenamefont {Barzel}\ and\ \citenamefont
  {Barab{\'a}si}(2013)}]{barzel2013universality}%
  \BibitemOpen
  \bibfield  {author} {\bibinfo {author} {\bibfnamefont {B.}~\bibnamefont
  {Barzel}}\ and\ \bibinfo {author} {\bibfnamefont {A.-L.}\ \bibnamefont
  {Barab{\'a}si}},\ }\href@noop {} {\bibfield  {journal} {\bibinfo  {journal}
  {Nature physics}\ }\textbf {\bibinfo {volume} {9}},\ \bibinfo {pages} {673}
  (\bibinfo {year} {2013})}\BibitemShut {NoStop}%
\bibitem [{\citenamefont {Bassett}\ and\ \citenamefont
  {Sporns}(2017)}]{bassett2017network}%
  \BibitemOpen
  \bibfield  {author} {\bibinfo {author} {\bibfnamefont {D.~S.}\ \bibnamefont
  {Bassett}}\ and\ \bibinfo {author} {\bibfnamefont {O.}~\bibnamefont
  {Sporns}},\ }\href@noop {} {\bibfield  {journal} {\bibinfo  {journal} {Nature
  neuroscience}\ }\textbf {\bibinfo {volume} {20}},\ \bibinfo {pages} {353}
  (\bibinfo {year} {2017})}\BibitemShut {NoStop}%
\bibitem [{\citenamefont {Kotsiantis}\ \emph {et~al.}(2007)\citenamefont
  {Kotsiantis}, \citenamefont {Zaharakis}, \citenamefont {Pintelas} \emph
  {et~al.}}]{kotsiantis2007supervised}%
  \BibitemOpen
  \bibfield  {author} {\bibinfo {author} {\bibfnamefont {S.~B.}\ \bibnamefont
  {Kotsiantis}}, \bibinfo {author} {\bibfnamefont {I.}~\bibnamefont
  {Zaharakis}}, \bibinfo {author} {\bibfnamefont {P.}~\bibnamefont {Pintelas}},
  \emph {et~al.},\ }\href@noop {} {\bibfield  {journal} {\bibinfo  {journal}
  {Emerging artificial intelligence applications in computer engineering}\
  }\textbf {\bibinfo {volume} {160}},\ \bibinfo {pages} {3} (\bibinfo {year}
  {2007})}\BibitemShut {NoStop}%
\bibitem [{\citenamefont {Mehta}\ \emph {et~al.}(2019)\citenamefont {Mehta},
  \citenamefont {Bukov}, \citenamefont {Wang}, \citenamefont {Day},
  \citenamefont {Richardson}, \citenamefont {Fisher},\ and\ \citenamefont
  {Schwab}}]{mehta2019high}%
  \BibitemOpen
  \bibfield  {author} {\bibinfo {author} {\bibfnamefont {P.}~\bibnamefont
  {Mehta}}, \bibinfo {author} {\bibfnamefont {M.}~\bibnamefont {Bukov}},
  \bibinfo {author} {\bibfnamefont {C.-H.}\ \bibnamefont {Wang}}, \bibinfo
  {author} {\bibfnamefont {A.~G.}\ \bibnamefont {Day}}, \bibinfo {author}
  {\bibfnamefont {C.}~\bibnamefont {Richardson}}, \bibinfo {author}
  {\bibfnamefont {C.~K.}\ \bibnamefont {Fisher}},\ and\ \bibinfo {author}
  {\bibfnamefont {D.~J.}\ \bibnamefont {Schwab}},\ }\href@noop {} {\bibfield
  {journal} {\bibinfo  {journal} {Physics reports}\ }\textbf {\bibinfo {volume}
  {810}},\ \bibinfo {pages} {1} (\bibinfo {year} {2019})}\BibitemShut {NoStop}%
\bibitem [{\citenamefont {Hopfield}(1982)}]{hopfield1982neural}%
  \BibitemOpen
  \bibfield  {author} {\bibinfo {author} {\bibfnamefont {J.~J.}\ \bibnamefont
  {Hopfield}},\ }\href@noop {} {\bibfield  {journal} {\bibinfo  {journal}
  {Proceedings of the national academy of sciences}\ }\textbf {\bibinfo
  {volume} {79}},\ \bibinfo {pages} {2554} (\bibinfo {year}
  {1982})}\BibitemShut {NoStop}%
\bibitem [{\citenamefont {Davidson}(2010)}]{davidson2010regulatory}%
  \BibitemOpen
  \bibfield  {author} {\bibinfo {author} {\bibfnamefont {E.~H.}\ \bibnamefont
  {Davidson}},\ }\href@noop {} {\emph {\bibinfo {title} {The regulatory genome:
  gene regulatory networks in development and evolution}}}\ (\bibinfo
  {publisher} {Elsevier},\ \bibinfo {year} {2010})\BibitemShut {NoStop}%
\bibitem [{\citenamefont {Mitchell}\ \emph {et~al.}(2016)\citenamefont
  {Mitchell}, \citenamefont {Tlusty},\ and\ \citenamefont
  {Leibler}}]{mitchell2016strain}%
  \BibitemOpen
  \bibfield  {author} {\bibinfo {author} {\bibfnamefont {M.~R.}\ \bibnamefont
  {Mitchell}}, \bibinfo {author} {\bibfnamefont {T.}~\bibnamefont {Tlusty}},\
  and\ \bibinfo {author} {\bibfnamefont {S.}~\bibnamefont {Leibler}},\
  }\href@noop {} {\bibfield  {journal} {\bibinfo  {journal} {Proceedings of the
  National Academy of Sciences}\ }\textbf {\bibinfo {volume} {113}},\ \bibinfo
  {pages} {E5847} (\bibinfo {year} {2016})}\BibitemShut {NoStop}%
\bibitem [{\citenamefont {Rocks}\ \emph {et~al.}(2017)\citenamefont {Rocks},
  \citenamefont {Pashine}, \citenamefont {Bischofberger}, \citenamefont
  {Goodrich}, \citenamefont {Liu},\ and\ \citenamefont
  {Nagel}}]{rocks2017designing}%
  \BibitemOpen
  \bibfield  {author} {\bibinfo {author} {\bibfnamefont {J.~W.}\ \bibnamefont
  {Rocks}}, \bibinfo {author} {\bibfnamefont {N.}~\bibnamefont {Pashine}},
  \bibinfo {author} {\bibfnamefont {I.}~\bibnamefont {Bischofberger}}, \bibinfo
  {author} {\bibfnamefont {C.~P.}\ \bibnamefont {Goodrich}}, \bibinfo {author}
  {\bibfnamefont {A.~J.}\ \bibnamefont {Liu}},\ and\ \bibinfo {author}
  {\bibfnamefont {S.~R.}\ \bibnamefont {Nagel}},\ }\href@noop {} {\bibfield
  {journal} {\bibinfo  {journal} {Proceedings of the National Academy of
  Sciences}\ }\textbf {\bibinfo {volume} {114}},\ \bibinfo {pages} {2520}
  (\bibinfo {year} {2017})}\BibitemShut {NoStop}%
\bibitem [{\citenamefont {Yan}\ \emph {et~al.}(2017)\citenamefont {Yan},
  \citenamefont {Ravasio}, \citenamefont {Brito},\ and\ \citenamefont
  {Wyart}}]{yan2017architecture}%
  \BibitemOpen
  \bibfield  {author} {\bibinfo {author} {\bibfnamefont {L.}~\bibnamefont
  {Yan}}, \bibinfo {author} {\bibfnamefont {R.}~\bibnamefont {Ravasio}},
  \bibinfo {author} {\bibfnamefont {C.}~\bibnamefont {Brito}},\ and\ \bibinfo
  {author} {\bibfnamefont {M.}~\bibnamefont {Wyart}},\ }\href@noop {}
  {\bibfield  {journal} {\bibinfo  {journal} {Proceedings of the National
  Academy of Sciences}\ }\textbf {\bibinfo {volume} {114}},\ \bibinfo {pages}
  {2526} (\bibinfo {year} {2017})}\BibitemShut {NoStop}%
\bibitem [{\citenamefont {Rocks}\ \emph {et~al.}(2019)\citenamefont {Rocks},
  \citenamefont {Ronellenfitsch}, \citenamefont {Liu}, \citenamefont {Nagel},\
  and\ \citenamefont {Katifori}}]{rocks2019limits}%
  \BibitemOpen
  \bibfield  {author} {\bibinfo {author} {\bibfnamefont {J.~W.}\ \bibnamefont
  {Rocks}}, \bibinfo {author} {\bibfnamefont {H.}~\bibnamefont
  {Ronellenfitsch}}, \bibinfo {author} {\bibfnamefont {A.~J.}\ \bibnamefont
  {Liu}}, \bibinfo {author} {\bibfnamefont {S.~R.}\ \bibnamefont {Nagel}},\
  and\ \bibinfo {author} {\bibfnamefont {E.}~\bibnamefont {Katifori}},\
  }\href@noop {} {\bibfield  {journal} {\bibinfo  {journal} {Proceedings of the
  National Academy of Sciences}\ }\textbf {\bibinfo {volume} {116}},\ \bibinfo
  {pages} {2506} (\bibinfo {year} {2019})}\BibitemShut {NoStop}%
\bibitem [{\citenamefont {Hexner}\ \emph
  {et~al.}(2020{\natexlab{a}})\citenamefont {Hexner}, \citenamefont {Liu},\
  and\ \citenamefont {Nagel}}]{hexner2020periodic}%
  \BibitemOpen
  \bibfield  {author} {\bibinfo {author} {\bibfnamefont {D.}~\bibnamefont
  {Hexner}}, \bibinfo {author} {\bibfnamefont {A.~J.}\ \bibnamefont {Liu}},\
  and\ \bibinfo {author} {\bibfnamefont {S.~R.}\ \bibnamefont {Nagel}},\
  }\href@noop {} {\bibfield  {journal} {\bibinfo  {journal} {Proceedings of the
  National Academy of Sciences}\ }\textbf {\bibinfo {volume} {117}},\ \bibinfo
  {pages} {31690} (\bibinfo {year} {2020}{\natexlab{a}})}\BibitemShut {NoStop}%
\bibitem [{\citenamefont {Schoning}(1999)}]{schoning1999probabilistic}%
  \BibitemOpen
  \bibfield  {author} {\bibinfo {author} {\bibfnamefont {T.}~\bibnamefont
  {Schoning}},\ }in\ \href@noop {} {\emph {\bibinfo {booktitle} {40th Annual
  Symposium on Foundations of Computer Science (Cat. No. 99CB37039)}}}\
  (\bibinfo {organization} {IEEE},\ \bibinfo {year} {1999})\ pp.\ \bibinfo
  {pages} {410--414}\BibitemShut {NoStop}%
\bibitem [{\citenamefont {Pashine}\ \emph {et~al.}(2019)\citenamefont
  {Pashine}, \citenamefont {Hexner}, \citenamefont {Liu},\ and\ \citenamefont
  {Nagel}}]{pashine2019directed}%
  \BibitemOpen
  \bibfield  {author} {\bibinfo {author} {\bibfnamefont {N.}~\bibnamefont
  {Pashine}}, \bibinfo {author} {\bibfnamefont {D.}~\bibnamefont {Hexner}},
  \bibinfo {author} {\bibfnamefont {A.~J.}\ \bibnamefont {Liu}},\ and\ \bibinfo
  {author} {\bibfnamefont {S.~R.}\ \bibnamefont {Nagel}},\ }\href@noop {}
  {\bibfield  {journal} {\bibinfo  {journal} {Science advances}\ }\textbf
  {\bibinfo {volume} {5}},\ \bibinfo {pages} {eaax4215} (\bibinfo {year}
  {2019})}\BibitemShut {NoStop}%
\bibitem [{\citenamefont {Scellier}\ and\ \citenamefont
  {Bengio}(2017)}]{scellier2017equilibrium}%
  \BibitemOpen
  \bibfield  {author} {\bibinfo {author} {\bibfnamefont {B.}~\bibnamefont
  {Scellier}}\ and\ \bibinfo {author} {\bibfnamefont {Y.}~\bibnamefont
  {Bengio}},\ }\href@noop {} {\bibfield  {journal} {\bibinfo  {journal}
  {Frontiers in computational neuroscience}\ }\textbf {\bibinfo {volume}
  {11}},\ \bibinfo {pages} {24} (\bibinfo {year} {2017})}\BibitemShut {NoStop}%
\bibitem [{\citenamefont {Stern}\ \emph {et~al.}(2020)\citenamefont {Stern},
  \citenamefont {Arinze}, \citenamefont {Perez}, \citenamefont {Palmer},\ and\
  \citenamefont {Murugan}}]{stern2020supervised2}%
  \BibitemOpen
  \bibfield  {author} {\bibinfo {author} {\bibfnamefont {M.}~\bibnamefont
  {Stern}}, \bibinfo {author} {\bibfnamefont {C.}~\bibnamefont {Arinze}},
  \bibinfo {author} {\bibfnamefont {L.}~\bibnamefont {Perez}}, \bibinfo
  {author} {\bibfnamefont {S.~E.}\ \bibnamefont {Palmer}},\ and\ \bibinfo
  {author} {\bibfnamefont {A.}~\bibnamefont {Murugan}},\ }\href@noop {}
  {\bibfield  {journal} {\bibinfo  {journal} {Proceedings of the National
  Academy of Sciences}\ }\textbf {\bibinfo {volume} {117}},\ \bibinfo {pages}
  {14843} (\bibinfo {year} {2020})}\BibitemShut {NoStop}%
\bibitem [{\citenamefont {Kendall}\ \emph {et~al.}(2020)\citenamefont
  {Kendall}, \citenamefont {Pantone}, \citenamefont {Manickavasagam},
  \citenamefont {Bengio},\ and\ \citenamefont
  {Scellier}}]{kendall2020training}%
  \BibitemOpen
  \bibfield  {author} {\bibinfo {author} {\bibfnamefont {J.}~\bibnamefont
  {Kendall}}, \bibinfo {author} {\bibfnamefont {R.}~\bibnamefont {Pantone}},
  \bibinfo {author} {\bibfnamefont {K.}~\bibnamefont {Manickavasagam}},
  \bibinfo {author} {\bibfnamefont {Y.}~\bibnamefont {Bengio}},\ and\ \bibinfo
  {author} {\bibfnamefont {B.}~\bibnamefont {Scellier}},\ }\href@noop {}
  {\bibfield  {journal} {\bibinfo  {journal} {arXiv preprint arXiv:2006.01981}\
  } (\bibinfo {year} {2020})}\BibitemShut {NoStop}%
\bibitem [{\citenamefont {Stern}\ \emph {et~al.}(2021)\citenamefont {Stern},
  \citenamefont {Hexner}, \citenamefont {Rocks},\ and\ \citenamefont
  {Liu}}]{stern2021supervised}%
  \BibitemOpen
  \bibfield  {author} {\bibinfo {author} {\bibfnamefont {M.}~\bibnamefont
  {Stern}}, \bibinfo {author} {\bibfnamefont {D.}~\bibnamefont {Hexner}},
  \bibinfo {author} {\bibfnamefont {J.~W.}\ \bibnamefont {Rocks}},\ and\
  \bibinfo {author} {\bibfnamefont {A.~J.}\ \bibnamefont {Liu}},\ }\href@noop
  {} {\bibfield  {journal} {\bibinfo  {journal} {Physical Review X}\ }\textbf
  {\bibinfo {volume} {11}},\ \bibinfo {pages} {021045} (\bibinfo {year}
  {2021})}\BibitemShut {NoStop}%
\bibitem [{\citenamefont {Noest}(1986)}]{noest1986new}%
  \BibitemOpen
  \bibfield  {author} {\bibinfo {author} {\bibfnamefont {A.~J.}\ \bibnamefont
  {Noest}},\ }\href@noop {} {\bibfield  {journal} {\bibinfo  {journal}
  {Physical review letters}\ }\textbf {\bibinfo {volume} {57}},\ \bibinfo
  {pages} {90} (\bibinfo {year} {1986})}\BibitemShut {NoStop}%
\bibitem [{\citenamefont {Moreira}\ and\ \citenamefont
  {Dickman}(1996)}]{moreira1996critical}%
  \BibitemOpen
  \bibfield  {author} {\bibinfo {author} {\bibfnamefont {A.~G.}\ \bibnamefont
  {Moreira}}\ and\ \bibinfo {author} {\bibfnamefont {R.}~\bibnamefont
  {Dickman}},\ }\href@noop {} {\bibfield  {journal} {\bibinfo  {journal}
  {Physical Review E}\ }\textbf {\bibinfo {volume} {54}},\ \bibinfo {pages}
  {R3090} (\bibinfo {year} {1996})}\BibitemShut {NoStop}%
\bibitem [{\citenamefont {Janssen}(1997)}]{janssen1997renormalized}%
  \BibitemOpen
  \bibfield  {author} {\bibinfo {author} {\bibfnamefont {H.}~\bibnamefont
  {Janssen}},\ }\href@noop {} {\bibfield  {journal} {\bibinfo  {journal}
  {Physical Review E}\ }\textbf {\bibinfo {volume} {55}},\ \bibinfo {pages}
  {6253} (\bibinfo {year} {1997})}\BibitemShut {NoStop}%
\bibitem [{\citenamefont {Dickman}\ and\ \citenamefont
  {Moreira}(1998)}]{dickman1998violation}%
  \BibitemOpen
  \bibfield  {author} {\bibinfo {author} {\bibfnamefont {R.}~\bibnamefont
  {Dickman}}\ and\ \bibinfo {author} {\bibfnamefont {A.~G.}\ \bibnamefont
  {Moreira}},\ }\href@noop {} {\bibfield  {journal} {\bibinfo  {journal}
  {Physical Review E}\ }\textbf {\bibinfo {volume} {57}},\ \bibinfo {pages}
  {1263} (\bibinfo {year} {1998})}\BibitemShut {NoStop}%
\bibitem [{\citenamefont {Vojta}(2006)}]{vojta2006rare}%
  \BibitemOpen
  \bibfield  {author} {\bibinfo {author} {\bibfnamefont {T.}~\bibnamefont
  {Vojta}},\ }\href@noop {} {\bibfield  {journal} {\bibinfo  {journal} {Journal
  of Physics A: Mathematical and General}\ }\textbf {\bibinfo {volume} {39}},\
  \bibinfo {pages} {R143} (\bibinfo {year} {2006})}\BibitemShut {NoStop}%
\bibitem [{\citenamefont {Hexner}(2021{\natexlab{a}})}]{hexner2021adaptable}%
  \BibitemOpen
  \bibfield  {author} {\bibinfo {author} {\bibfnamefont {D.}~\bibnamefont
  {Hexner}},\ }\href@noop {} {\bibfield  {journal} {\bibinfo  {journal} {arXiv
  preprint arXiv:2103.08235}\ } (\bibinfo {year}
  {2021}{\natexlab{a}})}\BibitemShut {NoStop}%
\bibitem [{\citenamefont {Maxwell}(1864)}]{maxwell1864calculation}%
  \BibitemOpen
  \bibfield  {author} {\bibinfo {author} {\bibfnamefont {J.~C.}\ \bibnamefont
  {Maxwell}},\ }\href@noop {} {\bibfield  {journal} {\bibinfo  {journal} {The
  London, Edinburgh, and Dublin Philosophical Magazine and Journal of Science}\
  }\textbf {\bibinfo {volume} {27}},\ \bibinfo {pages} {294} (\bibinfo {year}
  {1864})}\BibitemShut {NoStop}%
\bibitem [{\citenamefont {Calladine}(1978)}]{CALLADINE}%
  \BibitemOpen
  \bibfield  {author} {\bibinfo {author} {\bibfnamefont {C.}~\bibnamefont
  {Calladine}},\ }\href@noop {} {\bibfield  {journal} {\bibinfo  {journal}
  {International Journal of Solids and Structures}\ }\textbf {\bibinfo {volume}
  {14}},\ \bibinfo {pages} {161} (\bibinfo {year} {1978})}\BibitemShut
  {NoStop}%
\bibitem [{\citenamefont {Pellegrino}(1993)}]{PELLEGRINO1}%
  \BibitemOpen
  \bibfield  {author} {\bibinfo {author} {\bibfnamefont {S.}~\bibnamefont
  {Pellegrino}},\ }\href@noop {} {\bibfield  {journal} {\bibinfo  {journal}
  {International Journal of Solids and Structures}\ }\textbf {\bibinfo {volume}
  {30}},\ \bibinfo {pages} {3025 } (\bibinfo {year} {1993})}\BibitemShut
  {NoStop}%
\bibitem [{\citenamefont {Alexander}(1998)}]{alexander1998amorphous}%
  \BibitemOpen
  \bibfield  {author} {\bibinfo {author} {\bibfnamefont {S.}~\bibnamefont
  {Alexander}},\ }\href@noop {} {\bibfield  {journal} {\bibinfo  {journal}
  {Physics reports}\ }\textbf {\bibinfo {volume} {296}},\ \bibinfo {pages} {65}
  (\bibinfo {year} {1998})}\BibitemShut {NoStop}%
\bibitem [{\citenamefont {Hexner}\ \emph
  {et~al.}(2020{\natexlab{b}})\citenamefont {Hexner}, \citenamefont {Pashine},
  \citenamefont {Liu},\ and\ \citenamefont {Nagel}}]{hexner2020effect}%
  \BibitemOpen
  \bibfield  {author} {\bibinfo {author} {\bibfnamefont {D.}~\bibnamefont
  {Hexner}}, \bibinfo {author} {\bibfnamefont {N.}~\bibnamefont {Pashine}},
  \bibinfo {author} {\bibfnamefont {A.~J.}\ \bibnamefont {Liu}},\ and\ \bibinfo
  {author} {\bibfnamefont {S.~R.}\ \bibnamefont {Nagel}},\ }\href@noop {}
  {\bibfield  {journal} {\bibinfo  {journal} {Physical Review Research}\
  }\textbf {\bibinfo {volume} {2}},\ \bibinfo {pages} {043231} (\bibinfo {year}
  {2020}{\natexlab{b}})}\BibitemShut {NoStop}%
\bibitem [{\citenamefont {Maxwell}(1867)}]{maxwell1867iv}%
  \BibitemOpen
  \bibfield  {author} {\bibinfo {author} {\bibfnamefont {J.~C.}\ \bibnamefont
  {Maxwell}},\ }\href@noop {} {\bibfield  {journal} {\bibinfo  {journal}
  {Philosophical transactions of the Royal Society of London}\ ,\ \bibinfo
  {pages} {49}} (\bibinfo {year} {1867})}\BibitemShut {NoStop}%
\bibitem [{\citenamefont {Ellenbroek}\ \emph {et~al.}(2006)\citenamefont
  {Ellenbroek}, \citenamefont {Somfai}, \citenamefont {van Hecke},\ and\
  \citenamefont {van Saarloos}}]{ellenbroek2006critical}%
  \BibitemOpen
  \bibfield  {author} {\bibinfo {author} {\bibfnamefont {W.~G.}\ \bibnamefont
  {Ellenbroek}}, \bibinfo {author} {\bibfnamefont {E.}~\bibnamefont {Somfai}},
  \bibinfo {author} {\bibfnamefont {M.}~\bibnamefont {van Hecke}},\ and\
  \bibinfo {author} {\bibfnamefont {W.}~\bibnamefont {van Saarloos}},\
  }\href@noop {} {\bibfield  {journal} {\bibinfo  {journal} {Physical review
  letters}\ }\textbf {\bibinfo {volume} {97}},\ \bibinfo {pages} {258001}
  (\bibinfo {year} {2006})}\BibitemShut {NoStop}%
\bibitem [{\citenamefont {Lerner}\ \emph {et~al.}(2014)\citenamefont {Lerner},
  \citenamefont {DeGiuli}, \citenamefont {D{\"u}ring},\ and\ \citenamefont
  {Wyart}}]{lerner2014breakdown}%
  \BibitemOpen
  \bibfield  {author} {\bibinfo {author} {\bibfnamefont {E.}~\bibnamefont
  {Lerner}}, \bibinfo {author} {\bibfnamefont {E.}~\bibnamefont {DeGiuli}},
  \bibinfo {author} {\bibfnamefont {G.}~\bibnamefont {D{\"u}ring}},\ and\
  \bibinfo {author} {\bibfnamefont {M.}~\bibnamefont {Wyart}},\ }\href@noop {}
  {\bibfield  {journal} {\bibinfo  {journal} {Soft Matter}\ }\textbf {\bibinfo
  {volume} {10}},\ \bibinfo {pages} {5085} (\bibinfo {year}
  {2014})}\BibitemShut {NoStop}%
\bibitem [{\citenamefont {Vogel}(1921)}]{VFT1}%
  \BibitemOpen
  \bibfield  {author} {\bibinfo {author} {\bibfnamefont {H.}~\bibnamefont
  {Vogel}},\ }\href@noop {} {\bibfield  {journal} {\bibinfo  {journal}
  {Physikalische Zeitschrift}\ }\textbf {\bibinfo {volume} {22}},\ \bibinfo
  {pages} {645} (\bibinfo {year} {1921})}\BibitemShut {NoStop}%
\bibitem [{\citenamefont {Fulcher}(1925)}]{VFT2}%
  \BibitemOpen
  \bibfield  {author} {\bibinfo {author} {\bibfnamefont {G.~S.}\ \bibnamefont
  {Fulcher}},\ }\href@noop {} {\bibfield  {journal} {\bibinfo  {journal}
  {Journal of the American Ceramic Society}\ }\textbf {\bibinfo {volume} {8}},\
  \bibinfo {pages} {339} (\bibinfo {year} {1925})}\BibitemShut {NoStop}%
\bibitem [{\citenamefont {Tammann}\ and\ \citenamefont {Hesse}(1926)}]{VFT3}%
  \BibitemOpen
  \bibfield  {author} {\bibinfo {author} {\bibfnamefont {G.}~\bibnamefont
  {Tammann}}\ and\ \bibinfo {author} {\bibfnamefont {W.}~\bibnamefont
  {Hesse}},\ }\href@noop {} {\bibfield  {journal} {\bibinfo  {journal}
  {Zeitschrift f{\"u}r anorganische und allgemeine Chemie}\ }\textbf {\bibinfo
  {volume} {156}},\ \bibinfo {pages} {245} (\bibinfo {year}
  {1926})}\BibitemShut {NoStop}%
\bibitem [{\citenamefont {Grassberger}\ and\ \citenamefont {de~la
  Torre}(1979)}]{grassberger1979reggeon}%
  \BibitemOpen
  \bibfield  {author} {\bibinfo {author} {\bibfnamefont {P.}~\bibnamefont
  {Grassberger}}\ and\ \bibinfo {author} {\bibfnamefont {A.}~\bibnamefont
  {de~la Torre}},\ }\href@noop {} {\bibfield  {journal} {\bibinfo  {journal}
  {Annals of Physics}\ }\textbf {\bibinfo {volume} {122}},\ \bibinfo {pages}
  {373} (\bibinfo {year} {1979})}\BibitemShut {NoStop}%
\bibitem [{\citenamefont {Hinrichsen}(2000)}]{hinrichsen2000non}%
  \BibitemOpen
  \bibfield  {author} {\bibinfo {author} {\bibfnamefont {H.}~\bibnamefont
  {Hinrichsen}},\ }\href@noop {} {\bibfield  {journal} {\bibinfo  {journal}
  {Advances in physics}\ }\textbf {\bibinfo {volume} {49}},\ \bibinfo {pages}
  {815} (\bibinfo {year} {2000})}\BibitemShut {NoStop}%
\bibitem [{Note1()}]{Note1}%
  \BibitemOpen
  \bibinfo {note} {See supplementary information}\BibitemShut {NoStop}%
\bibitem [{\citenamefont {O'Hern}\ \emph {et~al.}(2003)\citenamefont {O'Hern},
  \citenamefont {Silbert}, \citenamefont {Liu},\ and\ \citenamefont
  {Nagel}}]{Ohern}%
  \BibitemOpen
  \bibfield  {author} {\bibinfo {author} {\bibfnamefont {C.~S.}\ \bibnamefont
  {O'Hern}}, \bibinfo {author} {\bibfnamefont {L.~E.}\ \bibnamefont {Silbert}},
  \bibinfo {author} {\bibfnamefont {A.~J.}\ \bibnamefont {Liu}},\ and\ \bibinfo
  {author} {\bibfnamefont {S.~R.}\ \bibnamefont {Nagel}},\ }\href@noop {}
  {\bibfield  {journal} {\bibinfo  {journal} {Phys. Rev. E}\ }\textbf {\bibinfo
  {volume} {68}},\ \bibinfo {pages} {011306} (\bibinfo {year}
  {2003})}\BibitemShut {NoStop}%
\bibitem [{Note2()}]{Note2}%
  \BibitemOpen
  \bibinfo {note} {See the supplementary information for additional
  details}\BibitemShut {NoStop}%
\bibitem [{\citenamefont {Hexner}(2021{\natexlab{b}})}]{hexner2020training}%
  \BibitemOpen
  \bibfield  {author} {\bibinfo {author} {\bibfnamefont {D.}~\bibnamefont
  {Hexner}},\ }\href@noop {} {\bibfield  {journal} {\bibinfo  {journal} {Soft
  Matter}\ }\textbf {\bibinfo {volume} {17}},\ \bibinfo {pages} {4407}
  (\bibinfo {year} {2021}{\natexlab{b}})}\BibitemShut {NoStop}%
\end{thebibliography}%

\end{document}